\documentclass[twocolumn,trackchanges, resetfootnote]{aastex701}
\usepackage{amsmath,amsfonts,amssymb}
\usepackage{bm}
\usepackage{graphics}
\usepackage{graphicx}
\usepackage{here} 
\usepackage{type1cm}
\usepackage{multirow}
\usepackage{float}
\usepackage{natbib}
\usepackage{booktabs} 
\usepackage[normalem]{ulem}  
\usepackage{soul}
\usepackage{xcolor}

\hypersetup{
unicode=true,
setpagesize=false,
bookmarksnumbered=true,
bookmarksopen=true,
colorlinks=true,
linkcolor=blue,
citecolor=blue,
urlcolor=blue
}

\def\Cone{\hbox{C\,{\scriptsize I}}}
\def\Ctwo{\hbox{C\,{\scriptsize {II}}}}

\shorttitle{First Determination of the Cosmic Microwave Background Radiation Temperature at $z\!=\!0.68$}
\shortauthors{Kotani et al.}
\graphicspath{{./}{figures/}}
\received{December 27, 2025}
\revised{Feburary 5, 2026}
\accepted{Feburary 24, 2026}
\submitjournal{ApJ}
\begin{document}

\title{First Determination of the Cosmic Microwave Background Radiation Temperature \\ at $z\!=\!0.68$ Using Molecular Absorption Lines}

\correspondingauthor{Tatsuya Kotani}
\email{sci.tatsu.729@keio.jp}

\author[orcid=0009-0006-9842-4830,gname=Tatsuya,sname=Kotani]{Tatsuya Kotani}
\affiliation{School of Fundamental Science and Technology, Graduate School of Science and Technology, Keio University, 3-14-1 Hiyoshi, Kohoku-ku, Yokohama, Kanagawa 223-8522, Japan}
\email{sci.tatsu.729@keio.jp}

\author[orcid=0000-0002-5566-0634,gname=Tomoharu,sname=Oka]{Tomoharu Oka} 
\affiliation{School of Fundamental Science and Technology, Graduate School of Science and Technology, Keio University, 3-14-1 Hiyoshi, Kohoku-ku, Yokohama, Kanagawa 223-8522, Japan}
\affiliation{Department of Physics, Faculty of Science and Technology, Keio University, 3-14-1 Hiyoshi, Kohoku-ku, Yokohama, Kanagawa 223-8522, Japan}
\email{tomo@phys.keio.ac.jp}

\author[orcid=0000-0003-2735-3239,gname=Rei,sname=Enokiya]{Rei Enokiya} 
\affiliation{National Astronomical Observatory of Japan, 2-21-1 Osawa, Mitaka, Tokyo 181-8588, Japan}
\email{rei.enokiya@nao.ac.jp}

\begin{abstract}  
  \noindent
  We analyzed millimeter-wave data toward the quasar B0218+357 observed with the Atacama Large Millimeter/submillimeter Array and obtained absorption spectra of the $J$=2--1 and $J$=3--2 rotational transitions of HCN, HCO$^{+}$, HNC, H$^{13}$CN, and H$^{13}$CO$^{+}$ at the cosmological redshift of $z\!=\!0.68$.
  For HCN, HCO$^{+}$, and HNC, we identified two distinct absorption components that are common to both transitions, whereas a single component was detected in the isotopologue spectra. In this paper, we accurately evaluate the excitation temperatures and their uncertainties from the absorption strengths of these components, and use them to determine the CMB temperature.
  Uncertainties in the continuum covering factor were propagated into the excitation temperature via Monte Carlo sampling. We further corrected the observed optical depths for biases due to column-density nonuniformity by assuming a lognormal column-density distribution. Under the assumption that the rotational levels are in radiative equilibrium with the cosmic microwave background (CMB), we derived excitation temperature profiles in the optically thin regime. Because the excitation of HCO$^+$ is biased by an additional velocity component and partial collisional excitation, this species was excluded from the final determination of the CMB temperature. From a weighted mean of the excitation temperatures obtained from HCN and HNC, we determined the CMB temperature at $z\!=\!0.68$ to be ${4.50\pm0.17\,\mathrm{K}}$. This constitutes the first measurement of the CMB temperature at $z\!=\!0.68$ based on a quasar absorption line system and represents the most precise determination at this redshift, highly consistent with the standard Big Bang cosmological model.
\end{abstract}

%% You will be asked to selected these concepts during the submission process but this old "keyword" functionality is maintained in case authors want to include these concepts in their preprints.
\keywords{\uat{Observational cosmology}{1146} --- \uat{Cosmic microwave background radiation}{322} --- \uat{Quasar absorption line spectroscopy}{1317} --- \uat{Interstellar medium}{847}}

\section{Introduction} \label{sec:intro}
The cosmic microwave background (CMB; \citealp{Penzias_1965, Dicke_1965}) is a nearly uniform, isotropic blackbody radiation field with a current temperature of $2.72548\pm0.00057\,\mathrm{K}$ \citep{Fixsen_2009}. 
It constitutes one of the strongest observational pillars of the Big Bang theory and is interpreted as a snapshot of the hot, dense, ionized Universe prior to the recombination epoch, approximately $3.8\times10^{5}$ years after the Big Bang.
The CMB temperature ($T_{\mathrm{CMB}}$) decreases with the age of the Universe. In the standard Big Bang cosmology, it is predicted to evolve with the cosmic adiabatic expansion as,
\begin{equation}\label{eq:tcmb}
  T_\mathrm{CMB}(z)=T_0\,(1+z),
\end{equation}
where $z$ is the cosmological redshift and $T_0$ is the present-day CMB temperature.

Equation~(\ref{eq:tcmb}) shows that the measurements of $T_{\mathrm{CMB}}(z)$ at multiple redshifts provide a direct test of cosmological models. 
Measurements of $T_{\mathrm{CMB}}$ at $z\!>\!0$ have been obtained by three principal methods. 
The first method is multifrequency thermal Sunyaev--Zel'dovich (SZ) observations \citep[{e.g.,}][]{Battistelli_2002, Horellou_2005, Luzzi_2009, Luzzi_2015, Hurier_2014, Saro_2014, deMartino_2015}, which have yielded a large number of measurements primarily at $z\!\lesssim\!1$.
The second involves analyses of H$_2$O absorption lines against the CMB. This technique currently provides the highest-redshift measurement at $z\!=\!6.34$ \citep{Riechers_2022}; however, no other such samples have been identified to date, and the measurement carries a relatively large uncertainty ($\sim$30\%).
The third employs excitation analyses of atomic and molecular absorption lines observed toward quasars. Measurements based on electronic transition lines of species such as CO, [\Cone] or [\Ctwo] probe relatively high redshifts ($2\lesssim{z}\lesssim3$), but their uncertainties are comparatively large ($\sim\!{10}$--50\%) \citep{Ge_1997, Srianand_2000, Molaro_2002, Cui_2005, Klimenko_2020}.
In contrast, rotational transition lines of highly polar molecules in intervening absorption systems---i.e., molecular clouds located in the foreground of the quasar---provide measurements at intermediate redshifts with much smaller fractional errors; for example, the measurement at $z\!=\!0.89$ achieves a relative uncertainty of $\sim$1--2\% \citep{Muller_2013, Kotani_2025}.
Although such systems are rare owing to the small molecular impact parameters required, they currently provide the most precise determinations of $T_\mathrm{CMB}$ at $z>0.5$.
A recent analysis of the $z\!=\!0.89$ absorber toward PKS1830--211 \citep{Kotani_2025} demonstrated that excitation analyses of highly polar molecules in diffuse molecular gas can yield high-precision measurements of $T_\mathrm{CMB}$, providing a methodological foundation for the present study.
To date, all measurements obtained via these methods remain consistent with the predictions of the standard model.
As illustrated in Figure~4 of \citet{Kotani_2025}, previous measurements of $T_\mathrm{CMB}(z)$ span a wide range of redshifts and methods, exhibiting sizable scatter---particularly at intermediate redshifts---despite being broadly consistent with the standard model prediction.

B0218+357 is a quasar at $z\!=\!0.944$ \citep{Cohen_2003}, and a nearly face-on spiral galaxy \citep{York_2005} is known to lie along its line of sight at $z=0.68466$ \citep{Browne_1993, Carilli_1993, WiklindCombes_1995}.
Due to the gravitational lensing produced by the intervening galaxy, the millimeter-wave image of B0218+357 is observed as two compact components on the sky, one located to the southwest and the other to the northeast, referred to as the A and B components, respectively \citep{Jethava_2007, Hada_2020}.
The A and B components are offset from the center of the spiral galaxy in the lens plane by approximately 0.29 arcsec (2.0 kpc) and 0.05 arcsec (0.4 kpc), respectively \citep{Hada_2020}.
Along the line of sight toward the A component at $z=0.68466$, a rich molecular absorption system arising in the spiral galaxy has been detected, first discovered by \citet{WiklindCombes_1995}. 
To date, numerous molecular species have been identified, including CO, CS, OH, C$_2$H, HCN, HCO$^+$, H$_2$O, H$_2$S, NH$_3$, H$_2$Cl$^+$, and H$_2$CO \citep{WiklindCombes_1995, MentenReid_1996, Combes_1997, CombesWiklind_1997, Gerin_1997, Kanekar_2003, Henkel_2005, Jethava_2007, Muller_2007, ZeigerDarling_2010, Kanekar_2011, Wallstrom_2016, Wallstrom_2019}. Isotopologues detected in this system include $^{13}$CO, C$^{34}$S, H$^{13}$CN, HC$^{15}$N, H$^{13}$CO$^+$, HC$^{18}$O$^+$, H$_{2}^{34}$S, and H$_2^{37}$Cl$^+$ \citep{Gerin_1997, Wallstrom_2016, Wallstrom_2019}.
This absorption system is one of few known intervening systems with rich molecular absorption features.

Both the A and B components of B0218+357 have not been previously used to measure the CMB temperature.
However, \citet{Jethava_2007} performed theoretical LVG calculations for the $z\!=\!0.68$ absorber toward the A component, targeting CS, HCN, HCO$^+$, HNC, and N$_2$H$^+$ and adopting physical conditions constrained by observations.
They pointed out that, under optically thin conditions, the excitation temperatures of the rotational transitions of these molecules are expected to provide a reliable probe of $T_\mathrm{CMB}(z\!=\!0.68)$.

In this work, we report the first measurement of $T_\mathrm{CMB}(z\!=\!0.68)$ using quasar absorption lines, derived from millimeter data obtained with the Atacama Large Millimeter/submillimeter Array (ALMA) toward the A component of B0218+357.
Using multiple highly polar molecules of HCN, HCO$^+$, HNC and their isotopologues, we carried out an analysis that carefully accounts for uncertainties in the continuum covering factor and temporal variability of the absorption strength, and nonuniformity of the column density (Sect.~\ref{sec:analy}). By combining the results obtained for the individual molecular species, we derived a new, high-precision measurement of the CMB temperature at $z\!=\!0.68$. 
This measurement, which probes the Universe roughly six billion years ago, places a tight constraint on deviations from the predictions of the standard Big Bang model (Sect.~\ref{sec:resdis}).

\section{Data Processing} \label{sec:data}
We retrieved the ALMA Band 3 and Band 4 data toward B0218+357 from the ALMA Science Archive\footnote{\url{https://almascience.eso.org/aq/}} (Project code 2016.1.00031.S; PI: S. Muller).
These observations contain the {\it J}=2--1 and {\it J}=3--2 absorption lines of HCN, HCO$^+$, HNC, H$^{13}$CN, and H$^{13}$CO$^+$, redshifted to $z\!=\!0.68466$.

Data calibration and reduction were carried out following the standard procedures in the Common Astronomy Software Applications (CASA) package\footnote{\url{http://casa.nrao.edu/}}. The Band 3 data were obtained on 2017 May 2 and July 8, while the Band 4 data were obtained on 2016 October 7 and 23. 
For the 2017 May 2 dataset, quasar J0510+1800 was used as the bandpass calibrator and J0423--0120 as the flux calibrator, whereas J0237+2848 and J0238+1636 were served as the bandpass and flux calibrators, respectively, for the remaining datasets.
Phase calibration was performed using quasar J0205+3212 for Band 3 and J0220+3241 for Band 4.
Subsequently, using the CASA task \texttt{mstransform}, we extracted the spectral windows containing the HCN, HCO$^+$, and HNC absorption lines with bandwidths of 0.23 GHz (0.47 GHz for the Band 4 data covering HNC), and the spectral windows containing the H$^{13}$CN and H$^{13}$CO$^{+}$ absorption lines with bandwidths of 0.12 GHz (Band 3) and 0.94 GHz (Band 4), respectively.

The spectra toward components A and B of B0218+357 were extracted using the Python-based package $\texttt{UVMULTIFIT}$ \citep{MartiVidal_2014} by directly fitting the visibilities for each component.
We adopted a two-point-source model and obtained the spectra through a two-step procedure: (1) determining the peak positions using only the continuum visibilities, and (2) fitting the visibilities in each spectral channel at those positions.
Although the two lensed images were not clearly resolved in the image plane, the simple and well-known two-point-source geometry of this system, with a separation of $\!\sim\!0{\overset{\prime\prime}{.}}3$ on the sky, enables spectral extraction via visibility fitting in the Fourier plane \citep[e.g.,][]{MartiVidal_2014, Wallstrom_2019}.
In the following analysis, we use the spectrum toward the A component, which exhibits a richer set of strong molecular absorption lines.

The frequency axis of the extracted spectra was converted into radial velocity using the relativistic Doppler relationship $V/c\!=\!\{1-({\nu}/{\nu_0})^2\}/\{1+({\nu}/{\nu_0})^2\}$, where $c$ and $\nu_0$ denote the speed of light and rest frequency, respectively.
The spectral axis was then regridded using the \texttt{mstransform} task to achieve the minimum velocity resolution specified in the ALMA archive, and the velocity reference frame was adopted as barycentric.
The redshift of the absorption system ($z\!=\!0.68466$) corresponds to a systemic velocity of $V_\mathrm{sys}\!=\!1.4357\times10^{5}$ $\mathrm{km\,s^{-1}}$.
When constructing the absorption spectra, velocities are expressed as offsets relative to this $V_\mathrm{sys}$.

\section{Analyses} \label{sec:analy}
\noindent
In this study, we follow the analysis methodology presented in \citet{Kotani_2025} to derive the optical depths and excitation temperatures from the molecular absorption line data, and to determine the CMB temperature.
Section~\ref{subsec:fundamentals} introduces the basis for deriving optical depth and excitation temperature.
Section~\ref{subsec:calculation} outlines the procedure for determining the CMB temperature from the excitation temperature.
Section~\ref{subsec:uncertainty} describes the sources of uncertainty in the excitation temperature, and Section~\ref{subsec:tau_cor} summarizes the corrections applied to the optical depths.

\subsection{Fundamentals}\label{subsec:fundamentals} 
We assumed that the line of sight toward the quasar is obscured by a foreground, homogeneous absorbing cloud.
The optical depth is computed from the continuum level ($I_\mathrm{c}$) and the absorption depth measured relative to it ($\Delta I$) as
\begin{equation}\label{eq:taudef}
  \tau_\nu=-\ln\left(1-\frac{\Delta{I}}{f_\mathrm{c}I_{\mathrm{c}}}\right),
\end{equation}
where $f_\mathrm{c}$ denotes the continuum covering factor.
With $f_\mathrm{c}$ specified, the optical depth profile can be derived from the absorption line profile using Equation~(\ref{eq:taudef}).

For pure rotational transitions of molecules, the optical depth of the $J\!\rightarrow\!J+1$ absorption line is related to the column densities of the lower and upper levels, $N_J$ and $N_{J+1}$, respectively.
Under the assumption that the molecular rotational transitions are characterized by a single excitation temperature ($T_{\rm ex}$) but are not necessarily thermalized \citep[weak LTE-conditions;][]{WiklindCombes_1995}, $N_{J+1}$ and $N_J$ follow a Boltzmann distribution determined by the $T_{\rm ex}$.
Therefore, the optical depth can be expressed in terms of the excitation temperature and the total molecular column density ($N$) as
\begin{eqnarray}\label{eq:tobesolved}
  &&\frac{8\pi\nu_{J+1,J}^3}{c^3A_{J+1,J}g_{J+1}}\int{\tau_{J+1,J}}\sqrt{\dfrac{1+{V}/{c}}{1-{V}/{c}}}\frac{1}{\left(1+{V}/{c}\right)^{2}}dV \quad\quad \nonumber \\
  && =\frac{N}{Z(T_{\mathrm{ex}})}\mathrm{exp}\left(-\frac{E_{J+1}}{k_\mathrm{B}T_{\mathrm{ex}}}\right)\left[\exp\left({\frac{h\nu_{J+1,J}}{k_\mathrm{B}T_{\mathrm{ex}}}}\right)-1\right],
\end{eqnarray}
where $\nu_{J+1,J}$ is the frequency of the $J\!\leftrightarrow\!J+1$ transition,
$A_{J+1,J}$ is the Einstein $A$ coefficient,
$g_J\!\equiv\!2J+1$ is the statistical weight of level $J$,
$\tau_{J+1,J}$ is the optical depth per unit velocity width,
$Z(T_{\mathrm{ex}})\!\equiv\!\sum{g_{J}\mathrm{exp}(-E_{J}/{k_\mathrm{B}T_{\mathrm{ex}} }) }$ is the partition function,
$E_J$ is the energy of level $J$, 
and $h$ and $k_\mathrm{B}$ are the Planck and Boltzmann constants, respectively.
$T_\mathrm{ex}$ and $N$ are obtained by fitting the right-hand side of Equation~(\ref{eq:tobesolved}) to the left-hand side. For each molecule, the two unknown parameters can be constrained by jointly using both the $J$=2--1 and $J$=3--2 transition lines. By redefining $N$ as the column density per velocity bin, Equation~(\ref{eq:tobesolved}) becomes applicable to individual velocity bins, allowing us to derive the excitation temperature profile from the velocity-integrated optical depth in each bin.

\subsection{$T_{\rm CMB}$ Calculations}\label{subsec:calculation} 
The population of molecular rotational energy levels in the interstellar medium is governed by both collisional and radiative processes.
In environments lacking local radiation fields and with extremely low densities of collision partners, such as diffuse gas, radiative coupling to the CMB becomes the dominant excitation mechanism. Because the relative contribution from collisional excitation scales with the number density of colliding particles, its importance can be assessed by comparing the gas density with the critical density.
Molecules with larger electric dipole moments have correspondingly higher critical densities; for example, the critical densities of the HCN, HNC, and HCO$^+$ {\it J}=1--0 transitions are of order $n_\mathrm{crit}({\rm H}_2)\!\sim\!{10^{5-6}}\,\mathrm{cm^{-3}}$ \citep{Shirley_2015}, especially in the optically thin limit, and even higher values are required for higher-$J$ transitions.
Therefore, in the absence of bright local radiation sources, rotational transitions of highly polar molecules embedded in low-density diffuse gas are expected to be dominated by radiative excitation by the CMB.

As a possible local radiation field surrounding the $z\!=\!0.68$ absorber toward the A component of B0218+357, thermal emission from dust may contribute.
The total hydrogen column density ($N_\mathrm{H}$) is a excellent tracer of the dust mass column density ($N_\mathrm{d}$) \citep[e.g.,][]{Bohlin_1978}, and their ratio, defined as $\delta_\mathrm{DG}\equiv{N_\mathrm{d}/(m_\mathrm{p}N_\mathrm{H})}$ \citep{Cheng_2025}, is known to be of order $\sim0.01$ in the Milky Way \citep{DraineHensley_2021}, where $m_\mathrm{p}$ is the proton mass.
Accordingly, the dust optical depth ($\tau_\mathrm{d}$) can be expressed using the mass absorption coefficient ($\kappa_\nu$) as
\begin{equation}
  \tau_\mathrm{d}=\kappa_\nu{N_\mathrm{d}}=\kappa_\nu{\delta_\mathrm{DG}m_\mathrm{p}N_\mathrm{H}}.
\end{equation}
Assuming that $N_\mathrm{H}$ toward the $z\!=\!0.68$ absorber is given by the sum of $N$($\mathrm{H}\,${\scriptsize I}) $\!=\!5\times10^{20}\,\mathrm{cm^{-2}}$ and $N({\rm H}_2)\!=$(0.1--2)$\times10^{22}\,{\rm cm}^{-2}$, as estimated by \citet{Henkel_2005}, the dust optical depth at the relevant frequencies is calculated to be $\tau_\mathrm{d}\sim10^{-7}$--$10^{-6}$.
Adopting this value of $\tau_\mathrm{d}$ together with a typical dust temperature at the solar position in the Milky Way, $T_\mathrm{d}=23\,\mathrm{K}$ \citep{Wright_1991}, we compare the intensities of the dust thermal emission ($I_\mathrm{d}$) and the CMB radiation ($I_\mathrm{CMB}$) at $z\!=\!0.68$ using
\begin{equation}
  \frac{I_\mathrm{d}}{I_\mathrm{CMB}}=\frac{B_\nu(T_\mathrm{d})(1-e^{-\tau_\mathrm{d}})}{B_\nu(T_\mathrm{CMB}(z))},
\end{equation}
where $B_\nu(T)$ is the Planck function.
The resulting ratio is found to be $\sim10^{-6}$--$10^{-5}$.
Thus, the dust radiation field is negligible compared to the CMB, and its contribution to the rotational excitation of the molecules considered in this study can be safely neglected in the $z\!=\!0.68$ absorption system.

The gas density and kinetic temperature of the $z\!=\!0.68$ absorber have been estimated to be $n(\mathrm{H_2})\!\sim\!{10^{2-4}\,\mathrm{cm^{-3}}}$ \citep{Henkel_2005, Jethava_2007, ZeigerDarling_2010} and $T_\mathrm{kin}\!=\!{55 \,\mathrm{K}}$ \citep{Henkel_2005}, respectively, suggesting that the absorbing medium consists predominantly of diffuse molecular gas \citep[e.g.,][]{Henkel_2005, Jethava_2007}.
Under these conditions, the rotational level populations of the highly polar molecules detected in this absorber, namely HCN, HCO$^+$, HNC, and their isotopologues, are in radiative equilibrium with the CMB.
Therefore, in this regime the relation
\begin{equation}\label{eq:tcmb=tex}
  T_\mathrm{CMB} = T_\mathrm{ex}
\end{equation}
holds. Indeed, the excitation analysis by \citet{Wallstrom_2016} assumes that collisional excitation is negligible compared to radiative excitation by the CMB. In addition, previous $T_\mathrm{CMB}$ measurements toward the SW component of PKS1830--211 at $z_\mathrm{abs}=0.89$ likewise relied on this relation, using absorption lines of highly polar molecules arising in diffuse molecular gas \citep{Wiklind_1996, Menten_1999, Kotani_2025}.
We also use this relation in our analysis.

\subsection{Uncertainty}\label{subsec:uncertainty}
\subsubsection{Continuum Covering Factor} \label{subsubsec:fc}
\setstcolor{red}
An incorrect assumption for $f_\mathrm{c}$ directly affects the inferred optical depth and can lead to a biased estimate of the excitation temperature.
For the A component, \citet{Muller_2007} estimated $f_\mathrm{c}\geq0.77$ from the saturated absorption bottom of the $\sim5\,\mathrm{km\,s^{-1}}$ component of the HCO$^+$ $J$=2--1 line.
However, a consensus value for $f_\mathrm{c}$ at millimeter wavelengths has not yet been established: \citet{Wallstrom_2016} adopted 0.8, whereas some study assumed $f_\mathrm{c}\!=\!1$ \citep{WiklindCombes_1995, Jethava_2007, ZeigerDarling_2010, Wallstrom_2019}.
Moreover, \citet{Jethava_2007} pointed out that $f_\mathrm{c}$ may vary with frequency.
These uncertainties in $f_\mathrm{c}$ represent a major source of error in the derived excitation temperature.

To rigorously account for both the indeterminacy and the possible frequency dependence of $f_\mathrm{c}$, we performed Monte Carlo calculations. 
We assumed that $f_\mathrm{c}$ follows a uniform distribution between 0.77 and 1.00. 
This choice reflects the weak physical constraints on $f_\mathrm{c}$ and thus adopts a maximally noninformative prior.
The absorption depth normalized by the continuum level, ${\Delta I/I_\mathrm{c}} = 1 - I_\mathrm{obs}$, is assumed to follow a normal distribution, with its mean and standard deviation given by the observed value and its uncertainty, respectively. Here, $I_\mathrm{obs}$ denotes the observed, continuum-normalized absorption spectrum at each velocity channel.
For each velocity bin and each transition, we drew random and independent samples of $f_\mathrm{c}$ and $\Delta I/I_\mathrm{c}$ from their respective distributions and computed the optical depth using Equation~(\ref{eq:taudef}).
Repeating this procedure 100,000 times yielded the probability distribution of the optical depth.
We adopted the median and the 1$\sigma$ confidence interval of this distribution as the observed optical depth ($\tau_\mathrm{obs}$) and its uncertainty.
In this way, the uncertainty in $f_\mathrm{c}$, including its possible time and frequency dependence, was fully propagated, together with the statistical uncertainty in $I_\mathrm{obs}$, into the final estimate of $\tau_\mathrm{obs}$.

\subsubsection{Time Variability}\label{subsubsec:time_var}
\noindent
Monitoring observations of the HCN and HCO$^{+}$ $J$=2--1 absorption toward B0218+357 with the IRAM Plateau de Bure Interferometer (PdBI) over a three-year period (2005--2008) revealed that 
the spectra do not exhibit absorption variations down to a few percent of the total continuum level of the combined A and B images \citep{Wallstrom_2016}. 
No millimeter monitoring observations have been carried out since this observation.
The Band 3 and Band 4 observations used in this study are separated by approximately nine months (from October 2016 to July 2017), but no flares, rapid millimeter-wavelength activity, or changes in the background quasar's continuum morphology have been reported during this interval.
On the basis of these considerations, we assume that no significant variability in the absorption strength occurred over this short period. 
Even if variability at the level of a few percent is assumed toward the A image alone in this analysis, the resulting change in the inferred CMB temperature is less than 0.1 K, which is well within the final statistical uncertainty.

\subsection{Correction for Nonuniformity}\label{subsec:tau_cor}
In Equation~(\ref{eq:taudef}), we assumed that the absorbing cloud has a uniform column density and partially covers the background quasar. 
In reality, however, the column density of the cloud is nonuniform. 
Regions with higher column density saturate the observed absorption depth $\Delta I/I_\mathrm{c}$, leading to an underestimation of $\tau_\nu$ and $N_J$. The effect of this column-density nonuniformity becomes more significant at larger optical depths and results in an overestimation of $T_{\rm ex}$.

\citet{Kotani_2025} corrected this bias by assuming that the column density follows a lognormal probability distribution function (N-PDF) and by computing a correction factor relating the observed optical depth ($\tau_{\rm obs}$) to the averaged actual optical depth ($\tau_{\rm real}$). The mean column density of the $z\!=\!0.68$ absorber toward the A component of B0218+357 is constrained to be $N({\rm H}_2)\!=$(0.1--2)$\times10^{22}\,{\rm cm}^{-2}$ \citep{Henkel_2005, Wallstrom_2016, Wallstrom_2019}. 
In this study, we adopted the N-PDF parameters derived by \citet{Ma_2020} for the Orion A giant molecular cloud, which has a comparable mean column density of $\sim 1.5\times10^{22}\,{\rm cm}^{-2}$ \citep{Ma_2020}, to compute the correction. 
This choice is further justified because lognormal N-PDFs are ubiquitously observed in Galactic molecular clouds, and observations also support lognormal-like distributions in low-density, extragalactic environments.
The resulting relation between $\tau_{\rm obs}$ and $\tau_{\rm real}$ is shown in Figure 1 of \citet{Kotani_2025}. We applied this nonuniformity correction to all data points in the absorption spectra.

\section{Results and Discussion} \label{sec:resdis}
\subsection{Optical Depth} \label{sec:opa}
\begin{figure*}[htbp]   \centering
  \includegraphics[width=\linewidth]{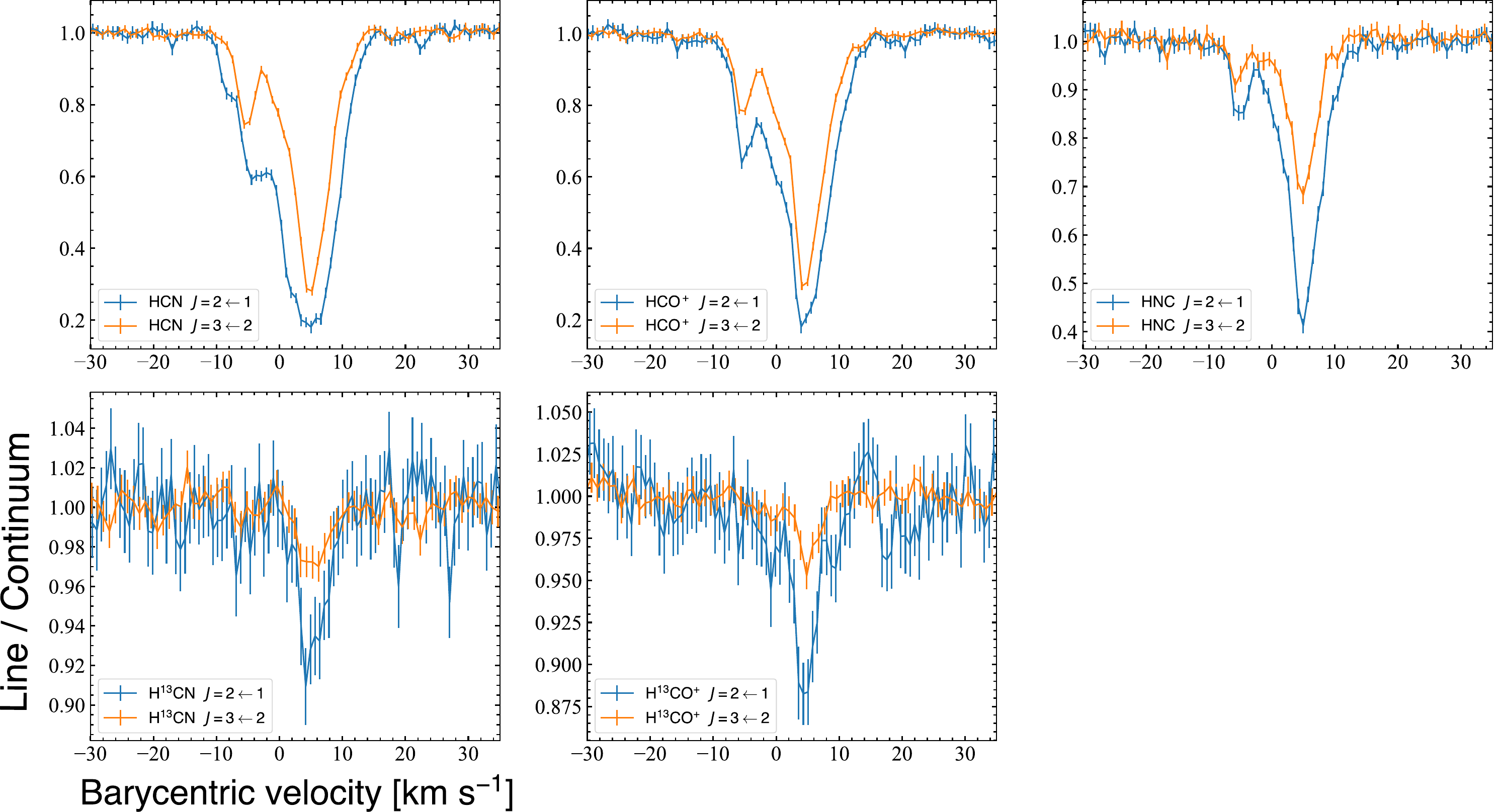}
  \caption{
  Continuum-normalized absorption spectra of the rotational transitions of HCN, HCO$^{+}$, HNC, H$^{13}$CN, and H$^{13}$CO$^{+}$ detected toward the A image of B0218+357. The velocities are shown as offsets from the systemic velocity, using the barycentric reference frame.
  }
  \label{fig:absprofile}
\end{figure*}

Figure \ref{fig:absprofile} presents the normalized absorption spectra of the HCN, HCO$^{+}$, HNC, H$^{13}$CN, and H$^{13}$CO$^{+}$ {\it J}=2--1 and {\it J}=3--2 transitions detected toward B0218+357 A. The HC$^{15}$N {\it J}=2--1 and {\it J}=3--2 transitions also fall within the observed spectral windows, but neither line is detected.
In the absorption profiles of HCN, HCO$^{+}$, and HNC, at least two components are commonly observed, centered at $V_{\rm bar}\!=\!+5\,\mathrm{km\,s^{-1}}$ and $-6\,\mathrm{km\,s^{-1}}$, respectively. 
These two components correspond to the c and a components reported by \citet{Muller_2007}, or to the first and third components described by \citet{Wallstrom_2016}.
For both HCN and HCO$^{+}$, the third absorption component, centered near $V_{\rm bar}\!=\! 0\,\mathrm{km\,s^{-1}}$, is present in the {\it J}=3--2 transition for HCN, whereas it is seen in both transitions for HCO$^{+}$. This feature is likely correspond to the b component identified by \citet{Muller_2007}.
The $+5\,\mathrm{km\,s^{-1}}$ component is also visible in the profiles of H$^{13}$CN and H$^{13}$CO$^{+}$, whereas the $-6\,\mathrm{km\,s^{-1}}$ component is not detected in either species.
Compared with the other molecules and transitions, the HCN {\it J}=2--1 line appears to be saturated at the bottom of the $+5\,\mathrm{km\,s^{-1}}$ component. This behavior is consistent with the PdBI results reported by \citet{Wallstrom_2016}. If we assume that the optical depth becomes infinite at the absorption minimum, Equation~(\ref{eq:taudef}) yields a lower limit of $f_{\rm c} \sim 0.8$, which is consistent with the constraint $f_{\rm c} \ge 0.77$.

The top panel of Figure~\ref{fig:Texprofile} shows the optical depth profiles of HCN, HCO$^+$, and HNC after accounting for the uncertainty in $f_{\rm c}$ and correcting for the bias introduced by the nonuniform column-density structure. Both HCN and HCO$^+$ are optically thick at the absorption peak of the $+5\,\mathrm{km\,s^{-1}}$ component in both transitions. For HNC, only the {\it J}=2--1 transition is optically thick at the peak of the $+5\,\mathrm{km\,s^{-1}}$ component. In contrast, the H$^{13}$CN and H$^{13}$CO$^{+}$ lines are optically thin in both transitions, with $\tau_{\rm max}\!\sim\!0.10$--0.15 for the {\it J}=2--1 lines and $\tau_{\rm max}\!\sim\!0.04$--0.05 for the {\it J}=3--2 lines.

\subsection{Excitation Temperature} \label{sec:Tex}
\noindent
In this study, following \citet{Kotani_2025}, we evaluated the optical depth on a per-velocity-channel basis for the high signal-to-noise ratio (S/N), optically thick HCN, HCO$^{+}$, and HNC lines. We then derived the excitation temperature by computing a weighted average of the resulting $T_\mathrm{ex}$ values.
For the low-S/N ($\sim3$), optically thin H$^{13}$CN and H$^{13}$CO$^{+}$ lines, we derived the excitation temperature using the optical depths integrated over the full velocity ranges exhibiting absorption.

\subsubsection{HCN, HCO$^{+}$, and HNC}\label{subsec:opticalythick}
All spectra of these molecules were resampled to match the velocity spacing of the {\it J}=3--2 data, which have the lowest spectral resolution ($\Delta V_\mathrm{bar}=0.909\,\mathrm{km\,s^{-1}}$).
$T_\mathrm{ex}$ and $N$ were determined by performing a weighted least-squares fit of Equation~(\ref{eq:tobesolved}) to the integrated optical depth calculated in each velocity bin. The initial value of $T_\mathrm{ex}$ was set to the standard Big Bang model prediction, $T_\mathrm{CMB}(z=0.68)=4.59\,\mathrm{K}$.
For the initial column densities, we adopted $7.8\times10^{13}\,\mathrm{cm^{-2}}$ and $4.5\times10^{13}\,\mathrm{cm^{-2}}$ for HCN and HCO$^{+}$, respectively, as reported by \citet{WiklindCombes_1995}.
For the HNC column density, an initial value of $2.6\times10^{13}\,\mathrm{cm^{-2}}$ was adopted.
Because no prior studies have analyzed the HNC line toward this source, we estimated the initial value using the abundance ratio $N(\mathrm{HNC})/N(\mathrm{HCN})$ measured in the PKS1830--211 absorber \citep{Muller_2011, Muller_2016}, which exhibits absorption from similar diffuse molecular gas, together with the HCN column density of \citet{WiklindCombes_1995}. 
These values were used to calculate the column density per velocity bin. 
To ensure that the derived excitation temperature reliably reflects the CMB temperature, the analysis was restricted to the velocity ranges where the optical depth is less than unity. In this regime, photon trapping and collisional excitation are expected to be negligible, and Equation~(\ref{eq:tcmb=tex}) is therefore applicable.

\begin{figure*}[htbp] 
  \centering
  \includegraphics[width=\linewidth]{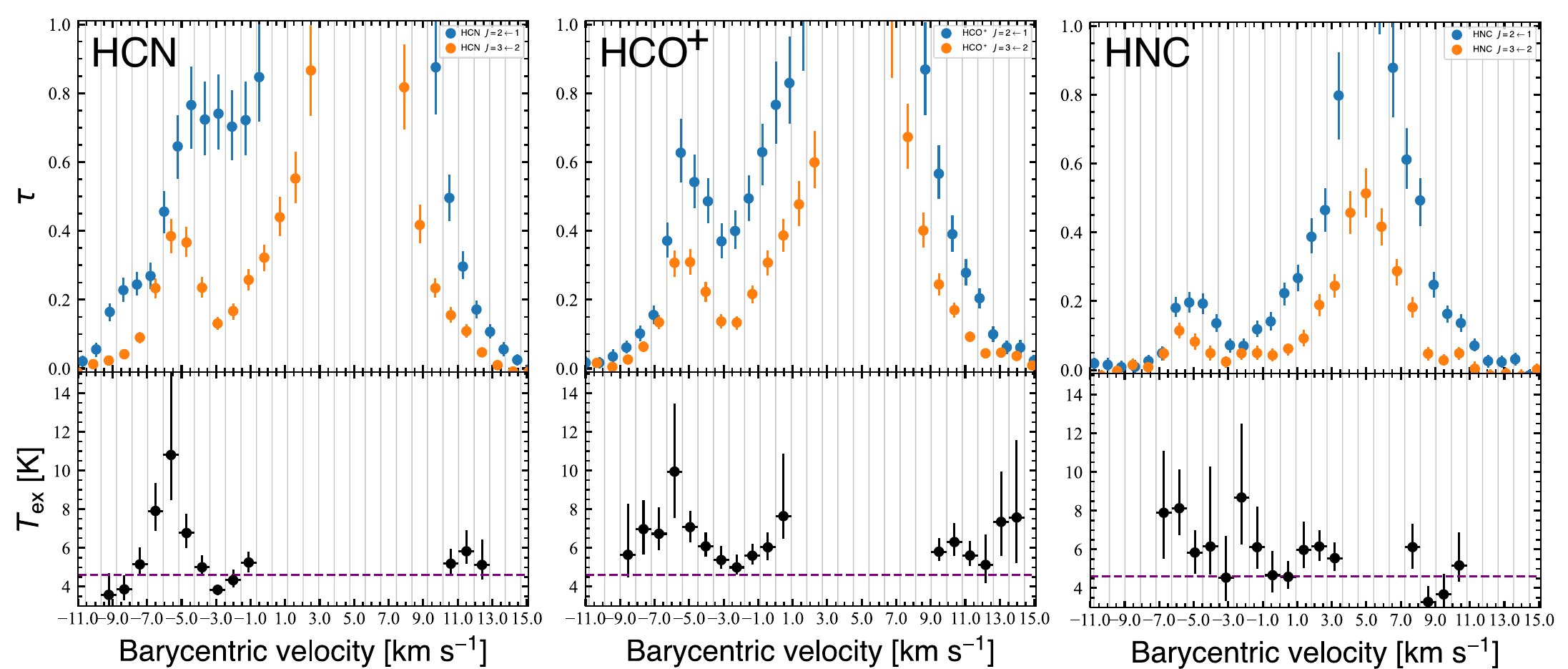}
  \caption{
    Profiles of HCN, HCO$^{+}$, and HNC obtained after accounting for the uncertainty in the continuum covering factor and the bias due to column-density inhomogeneity.
    (Top) Corrected optical depth profiles of the $J$=2--1 and $J$=3--2 transitions toward the A image of B0218+357, shown in blue and orange, respectively.
    Thin vertical lines indicate the boundaries of the velocity bins.
    (Bottom) Excitation temperature profiles calculated for velocity bins in which the optical depths of both the $J$=2--1 and $J$=3--2 transitions are smaller than unity. The purple dashed line marks the CMB temperature at $z\!=\!0.68$ predicted by the standard Big Bang model, $T_{\rm ex}\!=\!4.59\,\mathrm{K}$.
  }
  \label{fig:Texprofile}
\end{figure*}

The lower panels of Figure~\ref{fig:Texprofile} show the derived excitation temperature profiles.
The uncertainty in $T_\mathrm{ex}$ corresponds to the 68\% confidence interval of the Monte Carlo-derived distribution. Based on these profiles, we focused on the regions associated with the two velocity components. For the $+5\,\mathrm{km\,s^{-1}}$ component, the HCN $J$=2--1 absorption line appears to be saturated at its absorption center, allowing us to place a constraint on $f_\mathrm{c}$ that is consistent with previous studies. In contrast, the $-6\,\mathrm{km\,s^{-1}}$ component shows no evidence of saturation, preventing a meaningful constraint on $f_\mathrm{c}$ in that velocity range; consequently, the associated $T_\mathrm{ex}$ remains uncertain.
For this reason, we consider the excitation temperature of the $+5\,\mathrm{km\,s^{-1}}$ component to provide the most reliable estimate of the CMB temperature. 
We therefore computed, for each molecule, the weighted mean excitation temperature within the velocity range given in Table~\ref{table:Texresults}, using
\begin{equation}\label{meanTex}
  \left<T_{\rm ex} \right> \equiv \frac{\sum\left({{T_{\rm ex}}/{\sigma_{T_\mathrm{ex}}^2}}\right)}{\sum\left({{1}/{\sigma_{T_\mathrm{ex}}^2}}\right)}.
\end{equation}
Here, $\sigma_{T_\mathrm{ex}}$ represents the uncertainty in the excitation temperature at each velocity channel.
The resulting weighted mean values are summarized in Table~\ref{table:Texresults}.

\begin{deluxetable}{lccc}
  \tablecaption{Results of the molecular absorption-line analysis}\label{table:Texresults}
  \tablewidth{\columnwidth}
  \tablehead{
  \colhead{Molecule} &
  \colhead{Velocity range}\tablenotemark{a} &
  \colhead{$N_{\rm bin}$}\tablenotemark{b} &
  \colhead{$T_{\rm ex}$} \tablenotemark{c}\\
  &
  \colhead{($\mathrm{km\,s^{-1}}$)} &
  &
  \colhead{(K)}
  }
  \startdata
  HCN              & $-3.36$--$+12.9$  & 6 & $4.42^{+0.20}_{-0.20}$ \\
  HCO$^{+}$        & $-2.69$--$+12.6$  & 8 & $5.62^{+0.26}_{-0.25}$ \\
  HNC              & $-0.88$--$+10.8$  & 9 & $4.71^{+0.35}_{-0.32}$ \\
  H$^{13}$CN  & $-5.00$--$+15.0$  & \nodata & $4.93^{+0.61}_{-0.49}$ \\ 
  H$^{13}$CO$^{+}$ & $-5.00$--$+15.0$ & \nodata & $4.69^{+0.51}_{-0.40}$ \\ 
  \enddata
  \tablecomments{
    \tablenotemark{a} Velocity range corresponding to the $+5\,\mathrm{km\,s^{-1}}$ component. For HCN, HCO$^{+}$, and HNC, this indicates the velocity range used for the weighted mean, while for H$^{13}$CN and H$^{13}$CO$^{+}$ it corresponds to the velocity range used to integrate the optical depth.
    \tablenotemark{b} Number of velocity bins corresponding to the $+5\,\mathrm{km\,s^{-1}}$ component.
    \tablenotemark{c} For HCN, HCO$^{+}$, and HNC, this gives the weighted mean excitation temperature calculated from the integrated optical depth in each velocity bin, whereas for H$^{13}$CN and H$^{13}$CO$^{+}$ it gives the excitation temperature derived from the optical depth integrated over the full velocity range.
    }
\end{deluxetable}

To quantitatively assess the impact of collisional excitation, we performed RADEX \citep{vanderTak_2007} calculations adopting the physical conditions of the $z\!=\!0.68$ absorbing cloud ($T_\mathrm{kin}\!=\!55 \,\mathrm{K}$ and $n(\mathrm{H_2})\!=\!5\times10^3\, \mathrm{cm^{-3}}$) reported by \citet{Henkel_2005}. 
In this calculation, only the CMB was assumed as the radiation field. The line widths were set to approximately 17 $\mathrm{km\,s^{-1}}$, 17 $\mathrm{km\,s^{-1}}$, and 10 $\mathrm{km\,s^{-1}}$ for HCN, HCO$^{+}$, and HNC, respectively, based on the spectra reported in \citet{WiklindCombes_1995}.
The resulting excitation temperatures are approximately 4.7 K, 5.0 K, and 4.7 K for HCN, HCO$^{+}$, and HNC, respectively. 
For HCN, the $\langle T_{\rm ex} \rangle$ value listed in Table~\ref{table:Texresults} is lower than the corresponding value calculated with RADEX.
This suggests that, under the physical conditions of the $z\!=\!0.68$ absorber, collisional excitation plays a negligible role in populating the rotational levels of HCN.
For HNC, under the adopted physical conditions, the excitation temperature derived from RADEX is comparable to our measured value. However, if a higher gas density of $n(\mathrm{H_2})\!\sim\!10^4\,\mathrm{cm^{-3}}$ is assumed, as reported by \citet{ZeigerDarling_2010}, the RADEX calculations yield an excitation temperature that is higher than the value derived in our analysis.
These results indicate that, for both HCN and HNC in this absorber, the rotational excitation is governed predominantly by radiative excitation from the CMB rather than by collisional excitation. 

In contrast, even after accounting for uncertainties, the excitation temperature derived from the profile analysis for HCO$^{+}$ exceeds the RADEX calculations. 
Based on approximate estimates inferred from the results of \citet{Shirley_2015}, the critical densities of the $J$=2--1 and $J$=3--2 transitions of HCN, HCO$^{+}$, and HNC at $T_{\rm kin}=55\,\mathrm{K}$ are (2--6)$\times10^{6}\,{\rm cm^{-3}}$, (0.3--1)$\times10^{6}\,{\rm cm^{-3}}$, and (0.8--3)$\times10^{6}\,{\rm cm^{-3}}$, respectively, in the optically thin regime. The excitation temperatures exhibit an inverse dependence on the critical densities, in that species with higher critical densities have lower excitation temperatures. This indicates that the excitation-temperature behavior of HCO$^{+}$ found in our analysis is likely caused by the lower critical densities of the rotational levels of HCO$^{+}$ compared to those of HCN and HNC, which makes HCO$^{+}$ more susceptible to collisional excitation.
An alternative possibility is contamination from the third velocity component near $V_{\rm bar}\!=\! 0\,\mathrm{km\,s^{-1}}$, which appears to be in a higher excitation state and could bias the derived temperatures high. Indeed, this component is clearly confirmed in both the $J$=2--1 and $J$=3--2 transitions of HCO$^{+}$, and its excitation temperatures decrease systematically toward the line wings over the velocity range $V_{\rm bar}\!=\!-2.7$ to $+1.0\,\mathrm{km\,s^{-1}}$.
Therefore, the value of $\langle T_\mathrm{ex} \rangle$ derived for HCO$^{+}$ may not provide an accurate estimate of the CMB temperature.

For this reason, we regard the HCN and HNC results as the most reliable tracers of the CMB temperature in the $z\!=\!0.68$ absorber toward the A image of B0218+357. We then applied the weighted-mean procedure described by Equation~(\ref{meanTex}) to the excitation temperatures of these two species.
Under the relation expressed in Equation~(\ref{eq:tcmb=tex}), we obtain
\begin{equation}
  T_\mathrm{CMB}(z\!=\!0.68)=4.50\pm0.17\,\mathrm{K}.
\end{equation}
This value is in good agreement, within the uncertainties, with the prediction of the standard Big Bang model ($4.59~\mathrm{K}$). 
Compared with previous measurements at $z\!\sim\!0.68$ based on SZ observations, which yielded $4.78 \pm 0.28~\mathrm{K}$ at $z\!=\!0.676$ \citep{Hurier_2014} and $4.72^{+0.39}_{-0.27}~\mathrm{K}$ at $z\!=\!0.681$ \citep{Saro_2014}, our result provides a more precise determination.

\subsubsection{$^{13}$C isotopologues} \label{subsec:opticalythin}
The absorption spectra of H$^{13}$CN and H$^{13}$CO$^{+}$ have relatively low S/N.
We therefore fitted each optical depth profile with a single Gaussian component and computed the respective velocity-integrated optical depths over the range $-5\,\mathrm{km\,s^{-1}}$ to $+15\,\mathrm{km\,s^{-1}}$ for each transition of each molecule.
$T_\mathrm{ex}$ and $N$ were then determined by performing a weighted least-squares fit using Equation~(\ref{eq:tobesolved}).
The initial value of $T_\mathrm{ex}$ was set to the standard model prediction, as in Section~\ref{subsec:opticalythick}. 
The initial values of $N$ were adopted from \citet{Wallstrom_2016}, namely $2.9\times10^{12}\,\mathrm{cm^{-2}}$ for H$^{13}$CN and $2.0\times10^{12}\,\mathrm{cm^{-2}}$ for H$^{13}$CO$^{+}$.
The resulting excitation temperatures are listed in Table~\ref{table:Texresults}.
Although these uncertainties are larger than those derived for HCN and HNC, both values are consistent within uncertainties with the standard-model prediction of $4.59\,\mathrm{K}$.

For these molecules as well, we performed RADEX calculations adopting the same physical conditions as those applied in Section~\ref{subsec:opticalythick}, together with line widths of approximately 6.3 $\mathrm{km\,s^{-1}}$ and 4.5 $\mathrm{km\,s^{-1}}$ for H$^{13}$CN and H$^{13}$CO$^{+}$, respectively \citep{Wallstrom_2016}.
The resulting excitation temperatures are $\sim4.7\,\mathrm{K}$ and $\sim5.0\,\mathrm{K}$ for H$^{13}$CN and H$^{13}$CO$^{+}$, respectively.
The excitation temperature derived from the absorption profile is lower than that obtained from RADEX calculations for H$^{13}$CO$^{+}$, suggesting that its rotational levels are in radiative equilibrium with the CMB.
However, because the absorption is weak and the S/N is low for these isotopologues, we did not use them in the final estimate of the CMB temperature.

\subsection{Redshift Dependence of $T_\mathrm{CMB}$} \label{sec:cosmo}
\begin{figure}[htbp]
  \centering
  \includegraphics[width=\linewidth]{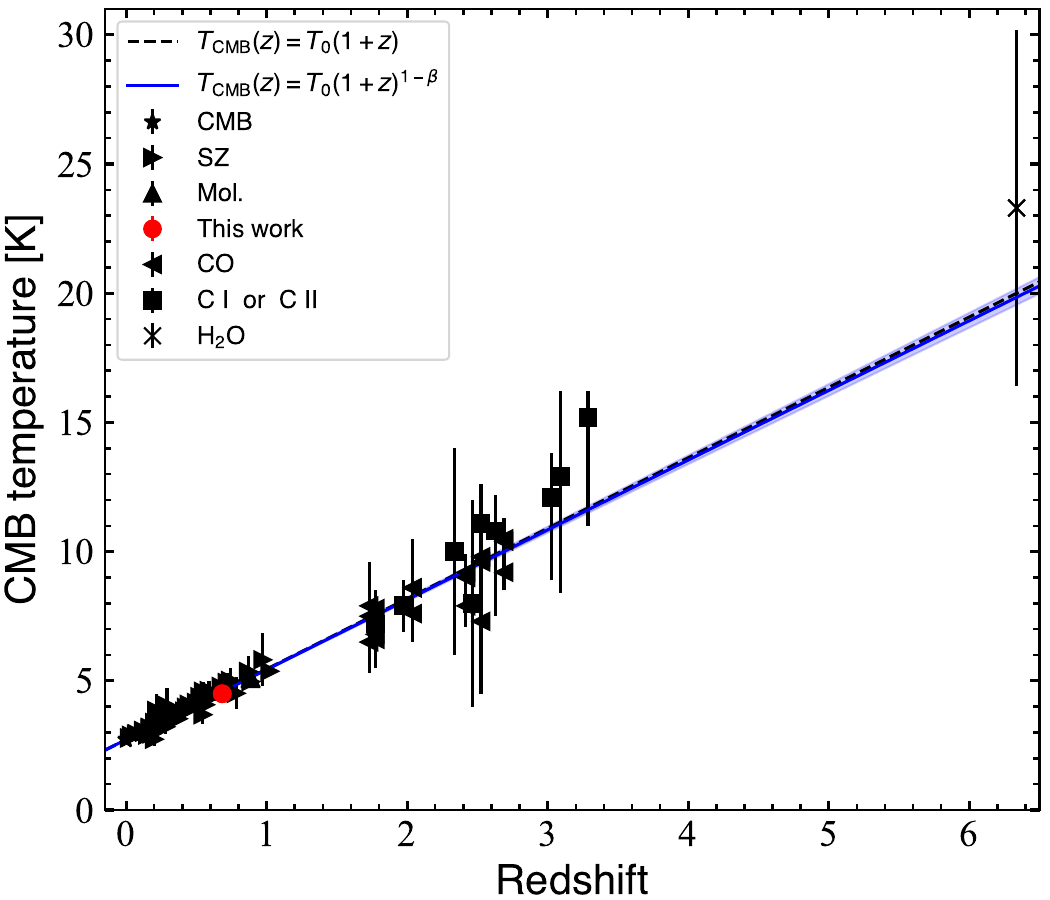}
   \caption{Plot of the measured CMB temperature ($T_\mathrm{CMB}$) versus redshift ($z$).  The values of previous measurements were taken from existing literature \citep[][and references therein]{Battistelli_2002, Luzzi_2009, Muller_2013, Klimenko_2020, Riechers_2022, Kotani_2025}.  
   A star indicates the measurement at $z\!=\!0$ \citep{Fixsen_2009}.  Each plot symbol represents a measurement method at $z>0$.  The black dashed line denotes the relationship expected from the standard model (Equation (\ref{eq:tcmb})).  The blue curve shows the best-fit result by the Equation (\ref{eq:btcmb}), $\beta=(3.9^{+7.4}_{-8.2})\times10^{-3}$, with the 1$\sigma$ uncertainty area (blue shadow).}
   \label{fig:z-Tcmb}
\end{figure}

\noindent
Figure~\ref{fig:z-Tcmb} presents a compilation of CMB temperature measurements as a function of cosmological redshift. In addition to the new measurement of $T_\mathrm{CMB}$ at $z\!=\!0.68$ obtained in this work, the figure includes approximately 80 previous measurements in the literature \citep[][and references therein]{Battistelli_2002, Luzzi_2009, Muller_2013, Klimenko_2020, Riechers_2022, Kotani_2025}.
We fitted these data using the phenomenological model proposed by \citet{Lima_2000},
\begin{equation}\label{eq:btcmb}
  T_\mathrm{CMB}(z)=T_0\,(1+z)^{1-\beta}, 
\end{equation}
and obtained ${\beta=(3.9^{+7.4}_{-8.2})\times10^{-3}}$.
This result indicates that the previously obtained measurements of $T_\mathrm{CMB}(z)$ are compatible with the standard Big Bang model and is consistent with recent constraints on $\beta$ reported by \citet{Riechers_2022} and \citet{Kotani_2025}. Our result thus places strong constraints on deviations from the standard cosmological model and reinforces the conclusions drawn from previous $T_\mathrm{CMB}(z)$ measurements.

\section{Summary and Future Prospects}
In this study, we analyzed molecular absorption lines detected in ALMA millimeter-wave observations toward the quasar B0218+357 and obtained the most precise measurement to date of the CMB temperature at $z\!=\!0.68$ independent of SZ observations.
We extracted absorption spectra of the {\it J}=2--1 and {\it J}=3--2 transitions of HCN, HCO$^{+}$, HNC, H$^{13}$CN, and H$^{13}$CO$^{+}$, and derived the excitation temperature $T_{\rm ex}$ and its associated uncertainty. 
The impact of the uncertainty in the continuum covering factor ($f_{\rm c}$) was evaluated using a Monte Carlo approach over the expected range of $f_{\rm c}$. 
Biases in the observed optical depths arising from inhomogeneities in the absorbing cloud were corrected by assuming a lognormal probability distribution for the column density.
Taking into account the physical conditions of the absorber, we computed the weighted mean of the excitation temperatures derived from HCN and HNC, yielding $T_\mathrm{CMB}(z\!=\!0.68)\!=\!4.50\pm0.17 \,\mathrm{K}$. 
This value is in good agreement with the prediction of the standard Big Bang model and places constraints on nonstandard cosmological scenarios. These constraints are consistent with those obtained in previous studies.
The excitation temperatures derived from H$^{13}$CN and H$^{13}$CO$^{+}$ are also consistent with the prediction of the standard Big Bang model, although their statistical significance is limited by the lower S/N of the data.

Obtaining high-S/N absorption lines of isotopologues of HCN, HCO$^{+}$, and HNC remains an important goal for achieving more robust estimates of the CMB temperature. 
Owing to their high critical densities, the excitation temperatures of their rotational levels are expected to serve as reliable tracers of the CMB temperature in the $z\!=\!0.68$ absorber toward B0218+357. Achieving this will require highly sensitive spectral-line observations.

For a more stringent test of the standard Big Bang cosmological model, it is also essential to increase the number of high-precision measurements of $T_\mathrm{CMB}$ at $z>1$. A natural extension of this work will be ALMA observations toward quasars hosting absorbers at $z\!=\!2\mbox{--}3$ along the line of sight, targeting highly polar molecular lines.
In addition, although they are relatively sensitive to collisional excitation, millimeter-wave fine-structure transitions of neutral carbon ([\Cone]) also constitute useful targets.
Compared to highly polar molecules, [\Cone] is more widely distributed in the interstellar medium, allowing for a larger sample of high-redshift absorption systems.
Indeed, analyses of [\Cone] absorption lines in the optical regime have already provided a number of measurements of the CMB temperature at high redshifts ($z\gtrsim2$) \citep[e.g.,][]{Klimenko_2020}. Future radio facilities, such as the Square Kilometre Array, and the upgraded ALMA, are expected to make significant contributions by increasing both the number and precision of $T_\mathrm{CMB}$ measurements at high redshift ($z\!>\!1$). 
This progress will enable increasingly sensitive tests with small deviations from the standard cosmological model.

%% Please use the acknowledgment and contribution environments. This will be anonomyized when the "anonymous" style option is used. 
\begin{acknowledgments}
  This paper makes use of the following ALMA data: ADS/JAO.ALMA\#2016.1.00031.S. ALMA is a partnership of ESO (representing its member states), NSF (USA) and NINS (Japan), together with NRC (Canada), NSTC and ASIAA (Taiwan), and KASI (Republic of Korea), in cooperation with the Republic of Chile. The Joint ALMA Observatory is operated by ESO, AUI/NRAO and NAOJ.
  T.K. acknowledges support from the Nordic ALMA Regional Centre (ARC) node based at Onsala Space Observatory.  
  The Nordic ARC node is funded through Swedish Research Council grant No 2019-00208.  
  T.O. acknowledges the support from JSPS Grant-in-Aid for Scientific Research (A) No. 20H00178.  
\end{acknowledgments}

%% To help institutions obtain information on the effectiveness of their  telescopes the AAS Journals has created a group of keywords for telescope 
%% Following the acknowledgments section, use the following syntax and the \facility{} or \facilities{} macros to list the keywords of facilities used in the research for the paper.  
\facility{ALMA.}

%% Similar to \facility{}, there is the optional \software command to allow authors a place to specify which programs were used during the creation of the manuscript. Authors should list each code and include either a citation or url to the code inside ()s when available.
\software{CASA \citep{CASATeam_2022}, Matplotlib \citep{Hunter_2007}, NumPy (\citealp{vanderWalt_2011, Harris_2020}), RADEX \citep{vanderTak_2007}, SciPy \citep{Virtanen_2020}, and UVMULTIFIT \citep{MartiVidal_2014}.}

%% For this sample we use BibTeX plus aasjournalv7.bst to generate the

% \bibliography{sample701}{}
% \bibliographystyle{aasjournalv7}

\end{document}